\documentclass[reprint,aps,prb,showpacs,floatfix,superscriptaddress,citeautoscript,a4paper,longbibliography]{revtex4-1}

\usepackage{graphicx}
\usepackage{epstopdf}
\usepackage{amsmath, amsfonts, amssymb, amsbsy, textcomp,epsfig,indentfirst,color}
\usepackage{overpic,subfig}
\usepackage{graphics}

\usepackage{hyperref}
\hypersetup{colorlinks=true, linkcolor=blue, citecolor=blue, urlcolor=blue, unicode=true}
\newcommand{\PbSnXSe}{Pb$_{1-x}$Sn$_{x}$Se}
\newcommand{\PbSnXTe}{Pb$_{1-x}$Sn$_{x}$Te}

\usepackage{color}

\begin{document}
\title{Quantum Spin Hall Effect in IV-VI Topological Crystalline Insulators}
\author{S.~Safaei}
\affiliation{Institute of Physics, Polish Academy of Sciences, Aleja Lotnik\'{o}w 32/46, 02-668 Warsaw, Poland}
\author{M.~Galicka}
\affiliation{Institute of Physics, Polish Academy of Sciences, Aleja Lotnik\'{o}w 32/46, 02-668 Warsaw, Poland}
\author{P.~Kacman}
\affiliation{Institute of Physics, Polish Academy of Sciences, Aleja Lotnik\'{o}w 32/46, 02-668 Warsaw, Poland}
\affiliation{Erna and Jacob Michael Visiting Professor at Braun Center for Submicron Research, Weizmann Institute of Science, Rehovot 76100, Israel}
\author{R.~Buczko}
\email{buczko@ifpan.edu.pl}
\affiliation{Institute of Physics, Polish Academy of Sciences, Aleja Lotnik\'{o}w 32/46, 02-668 Warsaw, Poland}

\date{\today}

%71.20.-b	Electron density of states and band structure of crystalline solids
%71.70.Ej	Spin-orbit coupling, Zeeman and Stark splitting, Jahn-Teller effect
%73.20.At	Surface states, band structure, electron density of states
%79.60.-i	Photoemission and photoelectron spectra

\begin{abstract}
{\bf{We envision that quantum spin Hall effect should be observed in $(111)$-oriented thin films of SnSe and SnTe topological crystalline insulators. Using a tight-binding
approach supported by first-principles calculations of the band structures we demonstrate that in these films the energy gaps in the two-dimensional band spectrum depend
in an oscillatory fashion on the layer thickness. These results as well as the calculated topological invariant indexes and edge state spin polarizations show that for
films $\thicksim$ 20--40 monolayers thick a two-dimensional topological insulator phase appears. In this range of thicknesses in both, SnSe and SnTe, (111)-oriented films edge
states with Dirac cones with opposite spin polarization in their two branches are obtained. While in the SnTe layers a single Dirac cone appears at the projection of the
$\mathbf{\overline{\mathit\Gamma}}$ point of the two-dimensional Brillouin zone, in the SnSe (111)-oriented layers three Dirac cones at $\mathbf{\overline{\mathit{M}}}$ points projections are predicted.}}
\end{abstract}
\maketitle

Topological insulators (TIs) and the quantum spin Hall (QSH) effect  attract significant interest for both fundamental and practical reasons. In the three-dimensional (3D)
and two-dimensional (2D) TIs the bulk insulating states are accompanied by metallic helical Dirac-like electronic states on the surface (edges) of the crystal. Due to the
time-reversal symmetry, these surface states are topologically protected against scattering at 180$^o$. In 2D TIs this means that the metallic edge states provide dissipationless, spin-polarized conduction channels. Such property makes these structures extremely interesting for low-power-consumption
electronics and spintronics. Thus, the quest for systems with topologically non-trivial edge states, which give rise to QSH effect, has become recently one of the
most important topics in condensed matter physics.

The QSH effect is the net result of two opposite polarized spin currents traveling in opposite directions along the edges of a 2D TI. It has been predicted by Kane and Mele in graphene~\cite{Kane1-PhysRevLett-2005}. In this material, however, it could occur only at unrealistically low temperatures, since the intrinsic spin-orbit coupling, which should
open a band gap at the Dirac points, is very weak. Subsequently, it has been proposed and confirmed experimentally that QSH state might arise in HgTe/HgCdTe quantum wells~\cite{Koenig-Science-2007}
and also in InAs/GaSb heterostructures~\cite{Knez_PRL-2011}. In these experiments the transition from insulating to conducting behavior has been observed.  In both
cases the conductivity was close to the $2e^2/h$ value expected for the two parallel quantum Hall channels. The spin polarized nature of the edge states has been shown much later and
only in the HgTe-based structure~\cite{Brune-NaturePhys-2012}. All these observations were made at very low temperatures, below 10 K, due to the small
energy gap in these systems.

In the search for new QSH structures many theoretical predictions have been made, in which various, sometimes exotic, chemical classes of 2D materials were considered.
These include slightly buckled honeycomb lattice of Si atoms (silicene)~\cite{Xing=Tao-An-APL-2013}, Bi bilayers on different substrates~\cite{Zhi-Quan-PRB-2013},
Ge(Bi$_x$Sb$_{1-x}$)$_2$Te$_4$ with various Bi concentrations~\cite{Singh-PRB88-2013}and functionalized ultrathin tin films (stanene)~\cite{Yong-Xu-PRL-2013}.
Recently, it has been also shown that 2D transition metal dichalcogenides (Mo or W) should form a new class of large-gap QSH insulators~\cite{Fu-2014}. Finally, a 2D TI with energy gap as large as 0.8 eV has been predicted for an overlayer of Bi grown on semiconductor Si(111) surface functionalized with one-third monolayer of halogen atoms~\cite{PNAS-2014}.
Despite all these theoretical
predictions, no stand-alone thin film or a thin film supported on a suitable substrate have been up to now experimentally demonstrated to harbor a QSH state.

In this article we describe a study of $(111)$-oriented thin films of well known IV-VI semiconductors SnSe and SnTe, which can be quite easily grown on a BaF$_2$ substrate by e.g. molecular beam epitaxy. Such layers have rock-salt crystal structure, despite that bulk SnSe has an orthorhombic structure.~\cite{private} Recently, it has been shown that these compounds are so-called topological crystalline insulators (TCIs). TCIs are
nontrivial insulators supporting surface Dirac fermions protected not by time-reversal but by crystal symmetry~\cite{Fu-PhysRevLett-2011}. The IV-VI semiconductors, i.e., SnSe and SnTe,
as well as PbTe and PbSe and the
substitutional solid solutions of Pb- and Sn-based chalcogenides were identified earlier as trivial insulators. This is because in these compounds the band inversion happens simultaneously at an even
number (four) of $L$ points in the Brillouin zone (BZ). The angle-resolved photoemission spectroscopy (ARPES) confirmed, however, that these materials belong to the TCI class.
It has been shown that metallic surface states exist on the $(001)$ surfaces of SnTe~\cite{Tanaka-NatPhys-2012} as well as \PbSnXSe~\cite{Dziawa-NatMater-2012} and \PbSnXTe~\cite{Xu-NatCommun-2012}.
These gapless $(001)$ surface states are supported by mirror symmetry.
It should be emphasized that due to the spin-orbit interactions the metallic surface states observed in real SnTe- or SnSe-based compounds have almost linear, Dirac-like dispersions.
It has been shown lately that in (001)-oriented thin films of TCIs hybridization leads to a new 2D TCI phase, which supports two pairs of spin-filtered edge states~\cite{Junwei-Liu-NatureMat-2014,Ozawa}. These edge states are protected solely by mirror symmetry  -- the topological phase is indexed by the mirror
Chern number $|N_{M}|=2$. In contrast, our calculations show that in ultrathin TCI films (SnTe and SnSe) grown not along $(001)$ but along $(111)$ crystallographic axis, the QSH state can be obtained. This is because the four $L$ point projections onto the 2D BZ of a (111)-oriented thin film are not equivalent. While the energy structures at the three $\overline{M}$ projections are the same by symmetry, the energy structure at the fourth projection at the $\overline{\mathit\Gamma}$ point is different. This results in a possibility that the band inversion takes place in an odd (either one or three) number of points.

\begin{figure}[hbt]
\centering
\includegraphics[width=0.45\textwidth]{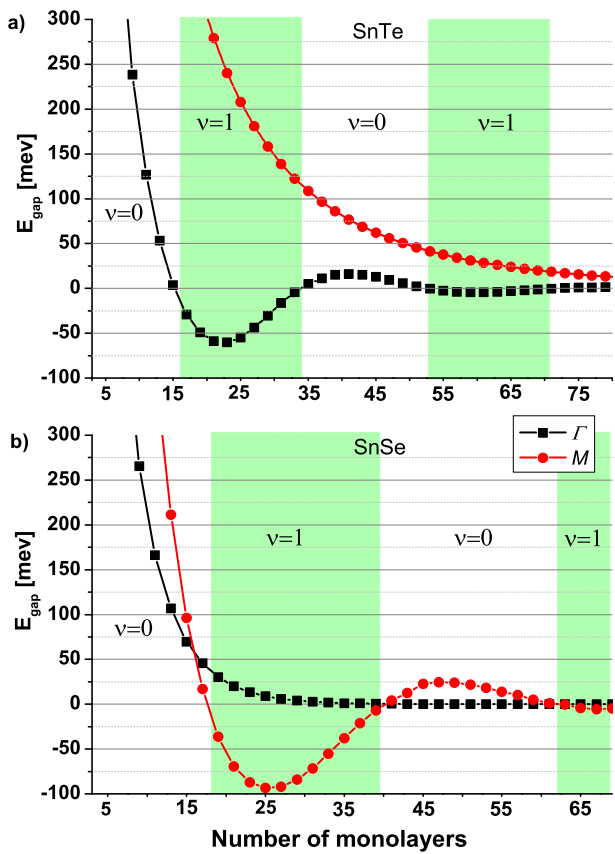}
\caption{{\bf Energy gaps in (111)-oriented films with an odd number of monolayers.} The dependence of the 2D states band gaps on the thickness of the (a) SnTe and (b) SnSe cation-terminated film. The green shadowed areas denote thicknesses, for which the value $\nu = 1$ has been obtained.}
  \label{fig:odd-gap-vs-thickness}
\end{figure}

A IV-VI crystal in rock-salt structure has \{111\} planes alternately composed of either cations or anions. Thus, (111)-oriented slabs consisting of an even number of monolayers
have one surface cation- and the other anion-terminated. In contrast, in slabs with an odd number of layers both surfaces are the same and the inversion symmetry is preserved. These two cases have to be distinguished.

\section{Odd number of monolayers}
For a (111)-oriented SnTe or SnSe thick film four single, topologically protected Dirac-cones in the four projections of the $L$-points onto the 2D BZ (one in $\overline{\mathit\Gamma}$
point and three in the $\overline{M}$ points) are obtained in the calculations. This is similar to what we have obtained before for a (111)-oriented bulk PbSnTe crystal~\cite{Safai-PRB-2013}.
For the anion-terminated surfaces, the bands are brought to contact forming anion Dirac cones, while in the other case the bands meet to form cation Dirac cones. As shown already in
Ref.~\onlinecite{Safai-PRB-2013,Liu-PRB-2013}, all $L$ points belong to the three \{110\} mirror planes of the (111)-surface and all Dirac points should be thus topologically protected. This predictions have been also confirmed experimentally for (Pb,Sn)Se~\cite{Creig} as well as for SnTe~\cite{Ando} case. While at the $\overline{\mathit\Gamma}$ point the Dirac-cone is isotropic, the band structure around $\overline{M}$ is strongly anisotropic, i.e., depends differently on the $k$ values along the $\overline{M}$ - $\overline{\mathit\Gamma}$ and $\overline{M}$ - $\overline{K}$ directions. This is due to different orientations of the constant energy ellipsoids around different $L$
points: the $L_1$ ellipsoid is projected on $\overline{\mathit\Gamma}$ along its long axis, whereas the ellipsoids for the other three $L$ points are inclined away from the projection direction. When the thickness of such TCI film decreases the gaps in the Dirac cones open, since the wave functions of the top and bottom surface states start to hybridize with each other.
Examples of band structures around $\overline{\mathit\Gamma}$ and around $\overline{M}$ points in the 2D BZ of (111)-oriented SnTe(SnSe) thin films are presented in Fig.~S1 (Fig.~S2) %\ref{fig:band111 slab}
in the Supplement. In Supplementary Section I the calculation method together with the obtained tight-binding parameters is also presented. Here in Fig.~\ref{fig:odd-gap-vs-thickness}
we present the 2D state energy gaps in ultrathin cation-terminated (111)-oriented SnTe (a) and SnSe (b) slabs as a function of the number (odd) of monolayers in the film.

As shown in Fig.~\ref{fig:odd-gap-vs-thickness}, in SnTe films the energy gap at the $\overline{M}$ point decreases monotonically with the increase of the film thickness. At the $\overline{\mathit\Gamma}$
point, however, damped oscillations of the energy gap with the number of monolayers can be observed. According to these results the first and biggest band gap inversion can be expected for SnTe films $\thicksim$ 3 - 6 nm
thick, i.e., consisting of 17-33 monolayers. In contrast, in SnSe thin films energy gap at $\overline{\mathit\Gamma}$ decreases monotonically, while the one at the $\overline{M}$ point oscillates with
the thickness of the film - in this case the first band gap inversion in $\overline{M}$ should occur for films with 19-39 monolayers, what corresponds to similar thicknesses of 3.3 - 6.7 nm.  Analogous results
are obtained for the anion-terminated layers. In this case, however, due to the smaller range of anion wave functions the inverted band gap starts to occur for slightly thinner layers (2.4 - 5.3 nm
for SnTe and 2.6 - 5.7 nm for SnSe), as shown in the Supplementary Section II. The damped oscillations of the band gap result from the $k^2$ corrections to the Dirac-like Hamiltonian of massive electron and have been already predicted for thin films of 3D TIs, like Bi$_2$Se$_3$ or Bi$_2$Te$_3$~\cite{Linder-2009,Lu-PRB81-2010}. In SnTe and SnSe $k^2$ terms are highly anisotropic.  The differences in the results obtained for films of these materials stem from the fact that in SnTe the $k^2$ term moves the energy gap from $L$ towards the $\mathit\Gamma$ point, whereas in SnSe towards the $W$ point (see Fig. S3 in the Supplement).

In analogy to results obtained in Ref.~\onlinecite{Ozawa} for (001)-oriented \PbSnXTe\ layers, the described above changes of the energy gaps are associated with a topological phase
transition. Our calculations show, however, that in the case of (111)-oriented SnTe and SnSe layers the topological phase changes from trivial insulator not to TCI, but to the 2D TI. The effect is similar to the transition to 2D TI predicted for Bi$_2$Se$_3$ or Bi$_2$Te$_3$ thin films~\cite{Lu-PRB81-2010}.
This result is obtained by calculations of the $Z_2$ invariant $\nu = 0,1$, which in 2D distinguishes between the trivial and QSH phase. For systems with inversion symmetry, i.e., for films consisting of an
odd number of monolayers, the parity of the occupied bands eigenstates at the four time-reversal momenta points in the BZ (one $\overline{\mathit\Gamma}$ and three $\overline{M}$) determines the $Z_2$ invariants~\cite{Fu-PRB76-2007}.
The green shadowed areas in Fig.~\ref{fig:odd-gap-vs-thickness} denote the SnTe and SnSe layer thicknesses, for which the value $\nu = 1$ has been obtained by the prescription given in
Ref.~\onlinecite{Fu-PRB76-2007}.
\begin{figure}[hbt]
\centering
\includegraphics[angle=-90,width=0.50\textwidth]{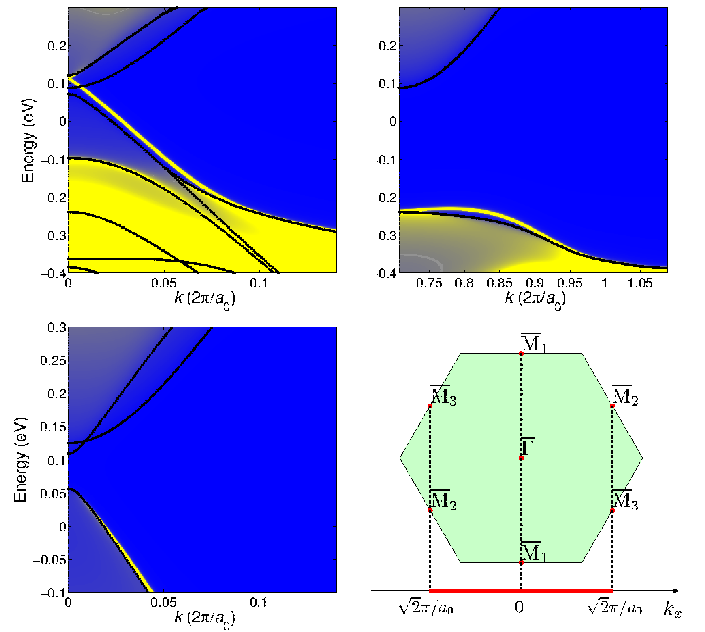}
\caption{{\bf Edge states in cation-terminated (111)-SnTe film.} The calculated spectral functions for a slab of 19 monolayers at the projection of (a) $\overline{\mathit\Gamma}$ and (b) $\overline{M_2}$ and $\overline{M_3}$ points onto the [${1\bar{1}0}$] edge. (c) The spectral function at the projection of $\overline{\mathit\Gamma}$ for 13 monolayers thick film. (d) The 2D BZ of (111)-oriented film (green hexagon) and 1D BZ of its [${1\bar{1}0}$] edge (red line). Local extrema of the 2D bands projected to the edge are here and in the following figures denoted by black dotted lines.}
  \label{fig:edge-odd-SnTe}
\end{figure}

\begin{figure}[hbt]
  \centering
  \includegraphics[angle=-90,width=0.50\textwidth]{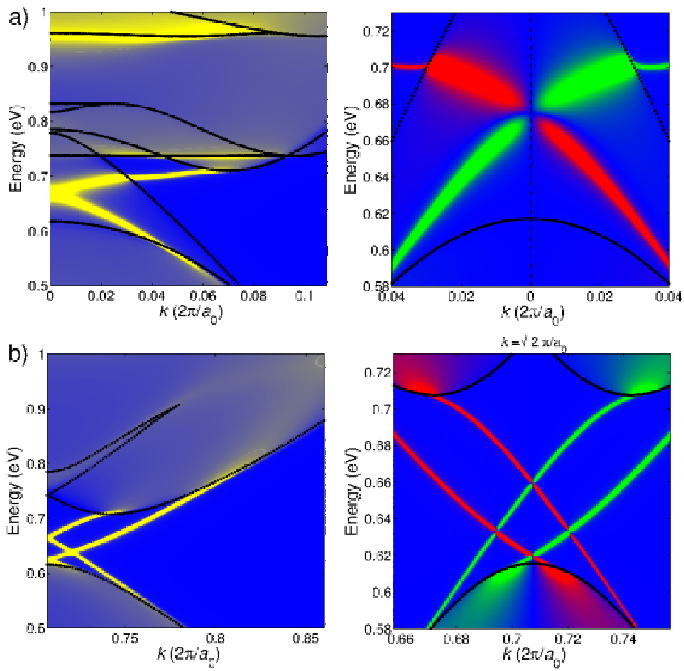}
  \caption{{\bf Edge states in 21-monolayers thick anion-terminated (111)-SnSe film}. The edge spectral functions in the vicinity of (a) $k_x=0$ (a) and (b) $k_x = {\sqrt{2}\pi}/{a_0}$ points of the 1D BZ. The corresponding spin densities are shown on the right. Red and green lines represent the spin-down and spin-up polarization, respectively.}
  \label{fig:edge-odd-SnSe}
\end{figure}

In Fig.~\ref{fig:edge-odd-SnTe}a and ~\ref{fig:edge-odd-SnTe}b we present the calculated [${1\bar{1}0}$] edge spectral functions of the (111)-SnTe
cation-terminated film consisting of 19 monolayers (i.e., with inverted band gap -- compare Fig.~\ref{fig:odd-gap-vs-thickness}a). The results for a slab consisting of 13 monolayers, where a trivial insulator phase is expected, is presented in Fig.~\ref{fig:edge-odd-SnTe}c. The 2D BZ zone of the (111)-oriented film and 1D BZ of the [${1\bar{1}0}$] edge are presented in Fig.~\ref{fig:edge-odd-SnTe}d. As shown in the latter, the $\overline{\mathit\Gamma}$ and $\overline{M}_1$ points project onto the edge at $k_x=0$, while $\overline{M}_2$ and $\overline{M}_3$ at $k_x = \pm {\sqrt{2}\pi}/{a_0}$. In agreement with the results of topological invariant calculations, shown above, Dirac crossing of the edge states appears in the band gap at the $\overline{\mathit\Gamma}$ projection for the 19-layers, but not for the 13-layers thick slab
(compare Fig.~\ref{fig:edge-odd-SnTe}a and ~\ref{fig:edge-odd-SnTe}c). At the projection of $\overline{M_2}$ and $\overline{M_3}$ (for $k_x = {\sqrt{2}\pi}/{a_0}$) the Dirac cones are not obtained for any SnTe layer thickness.
In contrast, in SnSe, as shown in Fig.~\ref{fig:edge-odd-SnSe}a and \ref{fig:edge-odd-SnSe}b for the
anion-terminated 21-monolayer thick slab, two Dirac crossings of the edge states appear inside the band gap at the $k_x = {\sqrt{2}\pi}/{a_0}$ point. As the third $\overline{M_1}$ point projects at $k_x=0$, another edge state with a Dirac
node appears at $k_x=0$. Due to the strong band overlapping, the latter is dispersed within the valence band.

\begin{figure}[hbt]
  \centering
  \includegraphics[width=0.45\textwidth]{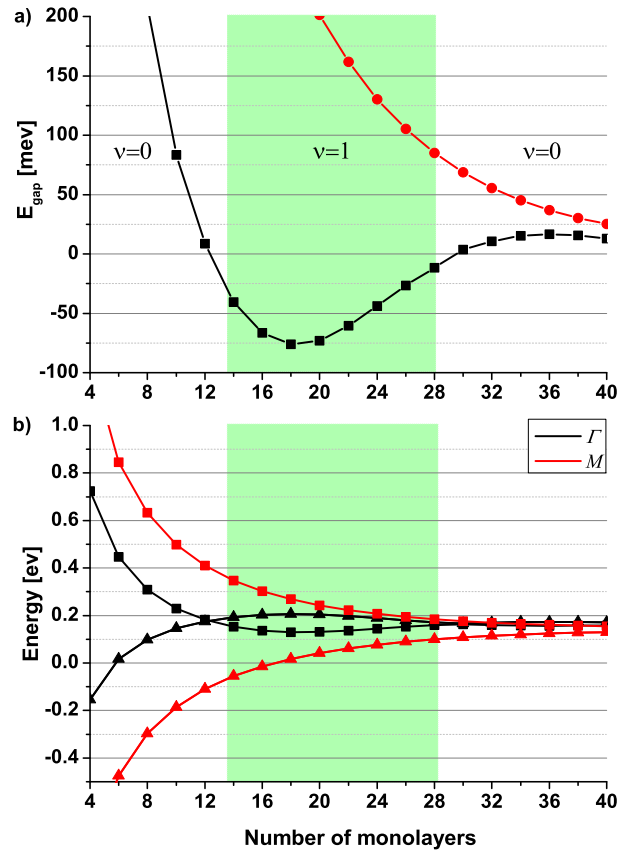}
  \caption{{\bf Energy gaps in SnTe film with an even number of monolayers.} (a) The dependence of the 2D state band gaps on the slab thickness. For thicknesses within the green shadowed area $\nu = 1$ was obtained. (b) The valence and conduction bands extrema in the vicinity of $\overline{\mathit\Gamma}$ (black) and $\overline{M}$ (red) vs the number of monolayers.}
  \label{fig:even-gap-vs-thickness}
\end{figure}
\begin{figure}[hbt]
  \centering
  \includegraphics[angle=-90,width=0.48\textwidth]{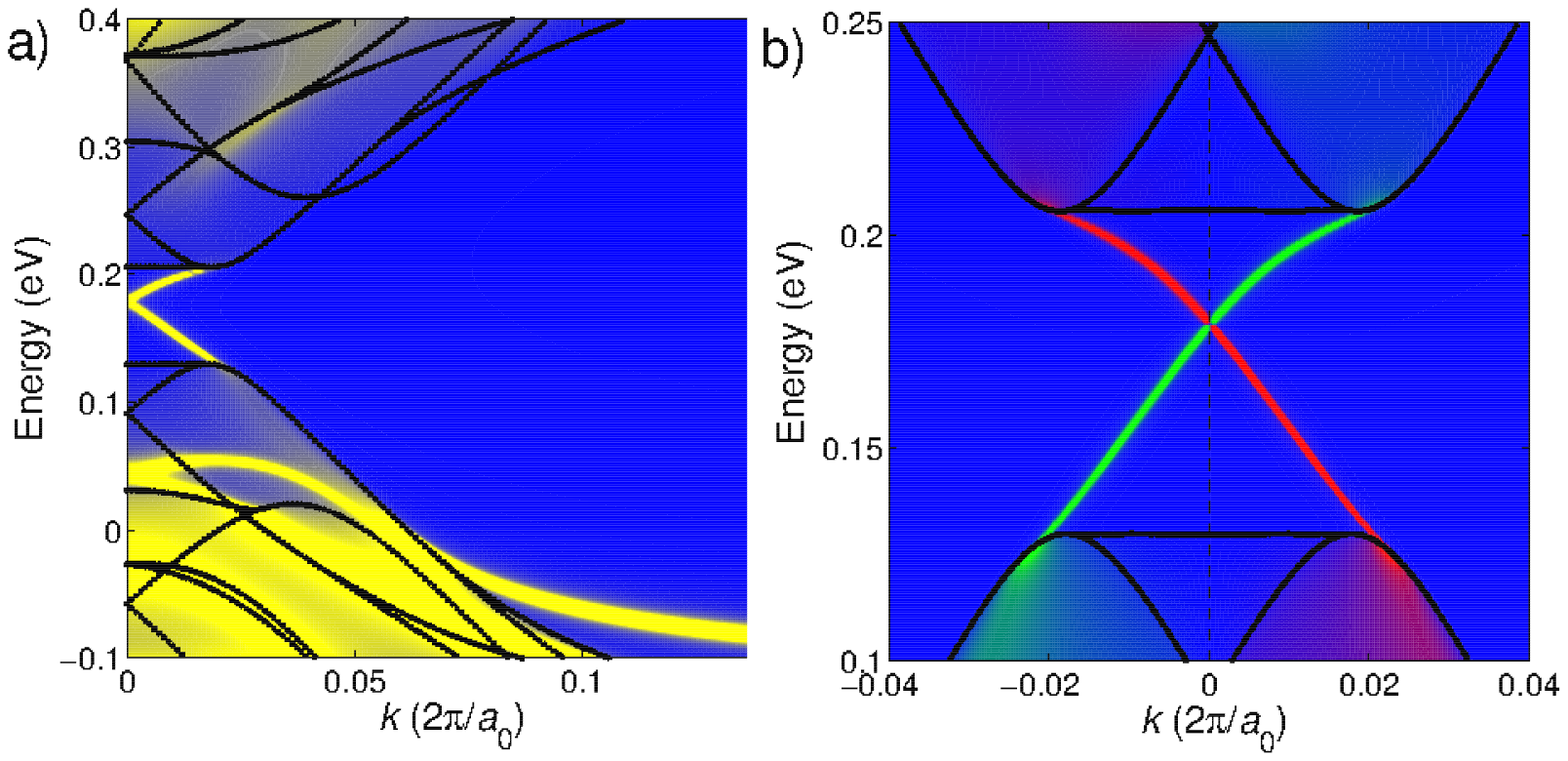}
  \caption{{\bf Edge states in 18 monolayers thick SnTe film}. (a) The edge spectral functions at the vicinity of $k_x=0$ point of the 1D BZ. (b) The corresponding spin density. The spin-down polarization is denoted by red, the spin-up by green color.}
  \label{fig:edge-even-SnTe}
\end{figure}

Finally, we have calculated the G$_{\uparrow}$ and G$_{\downarrow}$ contributions of the spin up and spin down (111)-projections in the edge spectral functions. The sign of the G$_{\uparrow}$ -
G$_{\downarrow}$ difference corresponds to spin polarization of the edge states. As expected for the 2D TI in the QSH state, the obtained edge states are spin polarized and the spin polarizations
in the two branches of the Dirac cones are always opposite. These spin polarizations in all studied cases are about 80\%, i.e.,
the contribution of the dominant spin configuration surpasses ca 8-10 times the opposite spin orientation.

\begin{figure}[hbt]
\centering
\includegraphics[angle=-90,width=0.48\textwidth]{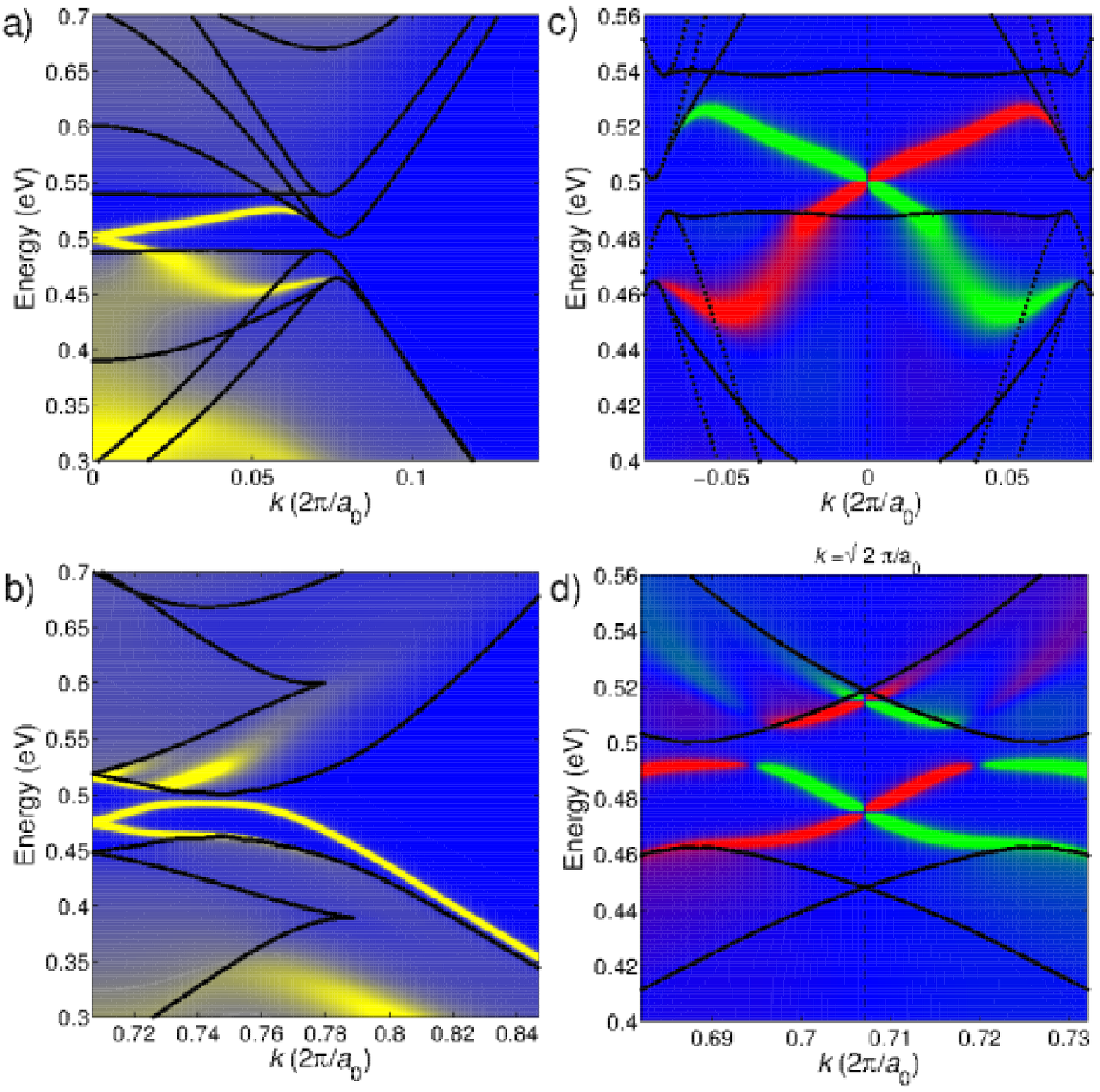}
\caption{{\bf Edge states in 20-monolayer thick SnSe film.} The edge spectral functions at $k_x=0$ are shown in (a), at the
$k_x = {\sqrt{2}\pi}/{a_0}$ point in (b). The spin polarizations of the edge states presented in (a) and (b) are shown in (c) and (d), respectively. Again, green color represents spin-up and red spin-down contribution to the edge spectral function.}
\label{fig:edge-even-SnSe}
\end{figure}

\section{Even number of monolayers}
Unlike the case of an odd number of layers, for (111)-oriented slabs consisting of an even number of monolayers there is no inversion symmetry in the system and the energy gaps are moved from the time-reversal invariant points (see Figs~S1(b) and S2(b) in the Supplementary Section I).
The gap inversion occurs now in two points related by the time reversal symmetry on two sides of each time reversal invariant $k$ point.
 It has been shown for inversion-asymmetric systems in general~\cite{Murakami-PRB-2007} that the gap closing can be accompanied by transition between the trivial insulator and the QSH phase when the gap inversion occurs in an odd number of such pairs. In SnSe films this takes place in three pairs in the vicinity of the three nonequivalent $\overline{M}$ points while in SnTe in one pair around the $\overline{\mathit\Gamma}$ point. Our calculations show that indeed for a given range of thicknesses the topological phase transition to the TI phase can be expected in these (111)-oriented films.

 If there is no inversion symmetry in the system, the $Z_2$ topological invariants can not be determined by the method utilizing the parity of the eigenfunctions, which we used for films with an odd number of monolayers. To calculate $Z_2$ for such layers we have adopted the method proposed by Fukai and Hatsugai~\cite{Fukai}. In this method the improvements to computing Chern numbers in a lattice  BZ enable the tight-binding calculations of $Z_2$ based on the formula derived by Fu and Kane in Ref.~\onlinecite{Fu_Kane_2006}.
The dependence of the band gap of SnTe (111)-oriented films consisting of an even number of monolayers on the layer thickness is shown in Fig.~\ref{fig:even-gap-vs-thickness}a. The thicknesses for which the value $\nu = 1$ has been obtained
are marked by the green shadow. As one can see in Fig.~\ref{fig:even-gap-vs-thickness}a  the QSH phase can be expected in SnTe films consisting of 14-28 monolayers  ($\thicksim$ 2.5 - 5.1 nm thick). For SnSe the QSH phase is obtained for 18-38 monolayers, i.e., for film thicknesses in the range 3.1 - 6.6 nm.
In the case of SnTe films with even number of layers the whole inverted energy gap for states in the vicinity of $\overline{\mathit\Gamma}$ is situated inside the large gap between conduction and valence bands close to the $\overline{M}$ point, as one can clearly see in Fig.~\ref{fig:even-gap-vs-thickness}b. The largest value of the inverted gap, about 75 meV, is obtained for the 18-20 monolayers thick SnTe slab.

The calculated spectral functions along the [${1\bar{1}0}$] edge for the (111)-oriented, 18-monolayers thick SnTe film show a clear Dirac node in the center of the band gap at the $k_x=0$ point.
This is shown in Fig.~\ref{fig:edge-even-SnTe}a.
%The existence of gapless helical edge modes with Dirac nodes situated in the time reversal invariant momenta of 1D BZ  follows directly from the formalism developed in Ref.~\onlinecite{Murakami-PRB-2007}.
The obtained results are in full agreement with the formalism developed in Ref.~\onlinecite{Murakami-PRB-2007}. We note that a similar result has been obtained for graphene~\cite{Kane1-PhysRevLett-2005} -- in this case one observes a single edge state with a Dirac
node at $\overline{\mathit\Gamma}$ (time-reversal symmetry point), despite the energy gaps appear in two points $K$ (in which the time-reversal is not preserved).

Fig.~\ref{fig:edge-even-SnSe} presents the edge spectral functions at the projections of $\overline{M}$ points at (a) $k_x=0$ and (b) $k_x = {\sqrt{2}\pi}/{a_0}$ point, calculated for the 20-monolayers thick SnSe slab. As
shown in the Supplementary Material, in this case the band gap at the $k_x=0$ is very small. Still, a pair of edge states can be seen near the top of the valence band. The spin polarization of this edge state is shown in Fig.~\ref{fig:edge-even-SnSe}c. At the $k_x = {\sqrt{2}\pi}/{a_0}$ point, however, the edge states which form
the two Dirac cones do not crossover the band gap to connect the valence and conduction bands. Instead, the upper and lower edge states repel each other and two anticrossings appear (see Fig.~\ref{fig:edge-even-SnSe}b). Interestingly, these
edge states are still spin polarized, as shown in Fig.~\ref{fig:edge-even-SnSe}d.

The calculations presented above show that in the free standing (111)-oriented SnSe and SnTe thin films there exists a range of thicknesses for which the 2D TI phase appears. The QSH phase is obtained for all studied films,
both with odd and also even number of monolayers. However, for all but the (111)-oriented SnTe films with an even number of monolayers an overlapping of bands in $\overline{\mathit\Gamma}$ and $\overline{M}$ diminishes
the final band gap. Hence, the edge states appear either against the background of the bands or within a very small energy gap (see Supplementary Section II). The (111)-oriented SnTe films with an even, close to 20, number of monolayers is thus the best candidate for observing the QSH effect in the IV-VI TCI. The QSH phase produced in this material system is highly robust, being even oblivious to the lack of inversion symmetry. The latter indicates that a suitable substrate or an overlayer will not destroy this 2D TI phase. Quite the contrary, using carefully selected substrate or overlay material (e.g., Pb$_{1-x}$Eu$_x$Se$_{1-y}$Te$_y$ quaternary alloy) would allow to tune the height of the energy barriers and biaxial strain in the structure~\cite{Springholz}. As shown already in Ref.~\onlinecite{Druppel} for SnTe (001) surface and for SnTe nanomembrane in Ref.~\onlinecite{Qian-NanoResearch-2014}, the uniaxial or biaxial strain can be used for manipulating with the energy gaps in the surface (or interface) states. The strain can be also used for solving the mentioned above problem of band overlapping, which diminishes the resulting energy gap (as it was shown  in Ref.~\onlinecite{Buczko} for solving a similar obstacle in PbTe/(Pb,Sn)Te heterostructures). Thus, these results may pave the road for experimental realization of QSH in SnTe and SnSe layers and their quantum well structures.

%Finally, it should be noted that all the above calculations have been performed for free standing layers, which can be difficult to obtain experimentally. Still, the result which predicts the QSH effect also in layers lacking inversion symmetry suggests that the presence of a substrate would not necessarily destroy the 2D TI phase.

%Also in Ref.~\onlinecite{Druppel} it has been shown that at the $(001)$ surface of SnTe a local gap at either all four or just two cones can be introduced via lattice deformations that break at least one of the underlying mirror symmetries.

\section{Acknowledgments}
We would like to thank Tomasz Story, Moty Heiblum, Junwei Liu and Pawel Potasz for valuable comments and suggestions. This work is supported by the Polish National Science Center (NCN) Grants No. 2011/03/B/ST3/02659 and 2013/11/B/ST3/03934. PK wishes to acknowledge the support of the Weizmann Institute Visiting Scientist Program. Calculations were partially carried out at the Academic Computer Center in Gdansk, using resources provided by Wroclaw Centre for Networking and Supercomputing (http://wcss.pl), grant No. 330  and at Interdisciplinary Centre for Mathematical and Computational Modelling (ICM) of the University of Warsaw under a grant No. G51-18.

\section{Contributions}
S.S. performed band-structure calculations, within both tight-binding and ab initio approaches, with contributions and supervision from R.B. M.G. performed the calculations of topological invariants and spin polarization. P.K. and R.B. performed the theoretical analysis and wrote the manuscript, with contributions from all authors. R.B. conceived and supervised the project. All correspondence should be addressed to R.B.

%\bibliography{referencesTCI}
%\bibliographystyle{apsrev4-1}
%\input{PRB_2013.bbl}

\end{document}